\documentclass[10.5,a4paper]{article}
\usepackage{amsmath, amssymb}
\usepackage{bm}
\usepackage{graphics}

\begin{document}

\begin{center}
{\LARGE \bf Bi-Local Field in Gravitational Shock Wave  \vspace{1mm}\\ Background} \vspace{5mm}\\

\noindent
N. Kanda\footnote{E-mail: \,nkanda@phys.cst.nihon-u.ac.jp}
 and S. Naka 
\footnote{E-mail:\, naka@phys.cst.nihon-u.ac.jp}
\vspace{2mm}\\

\noindent
{\small Department of Physics, College of Science and Technology Nihon University, Tokyo 101-8308, Japan}
\end{center}

\begin{minipage}{12cm}
{\small 
The particles with almost light velocity are able to be sources of the shock-wave gravity (SWG). Then, for ultra-high-energy particles, there exist two-body scatterings such that one particle is scattered from the gravitational background produced by another particle. Since the spacetime of SWG is closely related to a pp-wave solution of AdS-type background, this type of interaction is also interesting in AdS dual gauge theories. From those viewpoints, the scattering of point particles or strings by the SWG were studied. In this paper, we study the case of the bi-local models, which are simple relativistic bound systems having a close relation with specific modes of open strings. In particular, we analyze the bound-state effect on the scattering amplitudes, which describe the interaction between this model and SWG. }
\end{minipage}

\section{Introduction}

The analysis of particles scattered by shock-wave gravity provides an interesting way to study the scattering processes among ultra-high-energy particles as discussed by G. {}'t Hooft\cite{tHooft}. This is due to the observation by P. C. Aichelburg and R. S. Sexl\cite{Aichelburg} such that particles with almost light velocity in Minkowski spacetime are able to be sources of the SWG. The structure of such a background metric is also interesting from the viewpoint of AdS dual gauge theories, since the SWG has close relation with a pp-wave in string theories.

According to this line of approach, the scatterings of ultra-high energy projectiles by S.W.G. been studied in the cases of the particles\cite{Nastase}, the strings\cite{Amati}, and some kinds of fields\cite{Klimik}. The gravitational radiation from those collisions is also a matter of interest\cite{Veneziano}.

Now, the purpose of this paper is to study the scattering of bi-local models, the mechanical model of bi-local models of bi-local fields, by SWG. The bi-local fields were originally proposed by H. Yukawa in 1948 as an attempt of non-local fields\cite{Yukawa}. Through the development of those field theories, the bi-local models take on the characteristics of relativistic description of bound systems\cite{bi-local}. Nowadays, the bi-local models are sometimes understood in relation to a specific mode of an open-string model. If we consider a curved background spacetime, however, there arise essential differences between two models. The string model can be naturally embedded in a curved spacetime, since it is a one-dimensional continuum in spacetime. On the other hand, the embedding of the bi-local model in a curved spacetime is not trivial, since both ends of the bi-local models are supposed to interact by action at a distance.

In the next section, we study the embedding of bi-local models in a curved spacetime and in particular for the background spacetime of SWG to formulate the action of such a model. As in the case of the string models, the action of the bi-local model gives rise to constraints corresponding to a wave equation of the model and its subsidiary condition after the canonical quantization. Those constraints contain singularities which are characteristic to gravitational shockwave background; and, the canonical transformations, which remove singular factors in constraints, are also discussed.

In section 3, the scattering of the bi-local model by SWG is discussed.  Then, scattering amplitudes are evaluated for three processes under suitable approximation to getting meaningful results. Section 4 is devoted to the summary and discussions. 
In appendix A and B, brief reviews are given for the bi-local model in Minkowski spacetime and the Aichelburg-Sexl boost. We also add appendix B as an aid of the calculations in section 3.

\section{Bi-local model in curved spacetime}

\subsection{Bi-local model in gravitational shock-wave background}

We specify the shock-wave gravity produced by light-like particles by the metric
\footnote{In this paper, the metric in Minkowski spacetime is $\mbox{diag}(\eta_{\mu\nu})=(-+++)$. We also use the unit $\hbar=c=1$ except Appendix B. As for the coordinates $(x^{+},x^{-},x_\perp)$, the shockwave metric takes the form\\
$
 (g_{\mu\nu}) =\begin{pmatrix}
0 & -1 & 0 \\
-1 & f(x_\perp)\delta(x^{-}) & 0 \\
0 & 0 & {\bm 1}_\perp 
\end{pmatrix}; \mbox{that is},~~~
 (g^{\mu\nu})=\begin{pmatrix}
-f(x_\perp)\delta(x^{-}) & -1 & 0 \\
-1 & 0 & 0 \\
0 & 0 & {\bm 1}_\perp 
\end{pmatrix}. 
$ \\ }  
, the gravitational shock-wave background (Appendix B),
\begin{align}
 ds^2 &=g_{\mu\nu}(x)dx^{\mu}dx^{\nu}=-2dx^{+}dx^{-}+f(x_\perp)\delta(x^{-})d^2x^{-}+d^2x_{\perp},
\end{align}
where $x_{\perp}=(x^1,x^2),\, x^{\pm}=\frac{1}{\sqrt{2}}(x^{0}\pm x^{3})$ and
\begin{align}
 f(x_\perp)=f_0-2Q\log\left(\frac{r}{r_0}\right),~(r=\sqrt{x_\perp^2};\, Q\propto G:\mbox{grav. const.}).
\end{align} 
Here, $r_0$ and $f_0$ are constants associated with the normalization $ \left. f(x_\perp)\right|_{r=r_0}=f_0$. Since source particles of this S.W.G are considered to have the Planck scale energy $E_P$, $Q$ is proportional to $E_P^{-1}$ (\ref{Q}).

One natural way to construct the action of bi-local model in a curved spacetime is to extend (\ref{action}) in Minkowski spacetime in some way. A possible form of such an action is
\begin{align}
S =\int d\tau \frac{1}{2}\sum_{i=1}^2 \left\{ g_{\mu\nu}\frac{\dot{x}_{(i)}^\mu \dot{x}_{(i)}^\nu}{e_{(i)}}-V\left(x_{(1)},x_{(2)}\right)e_{(i)} \right\}, \label{action}
\end{align}
where $e_{(i)},(i=1,2)$ are einbeins in $\tau$ space. The $V(x_{(1)},x_{(2)})$ is a bi-scalar function representing the interaction of two particles with the same numerical value of $\tau$; further, we require
\footnote{
The center of mass coordinates and relative coordinates of the bi-local model are defined respectively by $X=\frac{1}{2}(x_{(1)}+x_{(2)})$ and $\bar{x}=x_{(1)}-x_{(2)}$.}
 $V\rightarrow \kappa^2\bar{x}^2+\omega,~(\bar{x}=x_{(1)}-x_{(2)},\,\omega=\mbox{const.})$ according as $g_{\mu\nu}\rightarrow \eta_{\mu\nu}$. As an interaction meet the requirement, we try to put $V(x_{(1)},x_{(2)})=2\kappa^2\sigma(x_{(1)},x_{(2)})+\omega$, where $\sigma(x_{(1)},x_{(2)})$ is the geodesic interval\cite{DeWitt} defined by
\begin{align}
 \sigma(x_{(1)},x_{(2)}) &=\frac{\Delta_{21}}{2}\int_{\gamma,\sigma_1}^{\sigma_2}d\sigma g_{\mu\nu}(x)\frac{\partial x^\mu}{\partial\sigma}\frac{\partial x^\nu}{\partial\sigma} ,~(\Delta_{21}=\sigma_2-\sigma_1),  \nonumber \\
 &=\frac{\Delta_{21}}{2}\int_{\gamma,\sigma_1}^{\sigma_2}d\sigma \left\{ -2x^{+\prime}x^{-\prime}+f(x_\perp)\delta(x^{-})(x^{-\prime})^2+x_\perp^{\prime 2} \right\} \label{geodesic-1}
\end{align}
with the notation \lq\lq prime\rq\rq as the derivative with respect to $\sigma$. Here, the $\kappa^{-1}$ represents the typical extension of the bi-local model, which is assumed to be very small compared with that of hadrons; in other words, the binding energy of this model is very strong. 

The $\sigma(x_{(1)},x_{(2)})$ is the one-half of the square of the distance along the geodesic $\gamma$ between $x_{(1)}=x(\sigma_1)$ and $x_{(2)}=x(\sigma_2)$. The right-hand side of (\ref{geodesic-1}) implies that the geodesic $\gamma$ is determined as a function of $\sigma$ with a constant $\tau$, the time-ordering parameter of particles attached to both ends of $\gamma$. Namely, the geodesic equations out of (\ref{geodesic-1}) include $\sigma$ derivatives only.

Now, the variation of (\ref{geodesic-1}) with respect to $x^{+}$ leads to $x^{-\prime\prime}=0$; and so, $x^{-}$ is the straight line $x^{-}(\sigma)=-\frac{\bar{x}^{-}}{\Delta_{21}}(\sigma-\sigma_1)+x_{(1)}^{-}$; that is, $x^{-\prime}=-\frac{1}{\Delta_{21}}\bar{x}^{-}$. Substituting this result for (\ref{geodesic-1}), we can write
\begin{align}
 \sigma(x_{(1)},x_{(2)}) =-\bar{x}^{+}\bar{x}^{-}+\frac{\Delta_{21}}{2}\int_{\gamma,\sigma_1}^{\sigma_2}d\sigma \left\{f(x_\perp)\delta(x^{-})(x^{-\prime})^2+x_\perp^{\prime 2} \right\}. 
\label{geodesic-2}
\end{align}
In order to make minimize the second term in the right-hand side of (\ref{geodesic-2}), it is sufficient to take the variation with respect to $x_\perp$, since the $x^{-}$ is already a definite function of $\sigma$. The result is $ x_\perp^{\prime\prime}=\frac{1}{2}(\partial_\perp f)_0\delta(x^{-})(x^{-\prime})^2=-Q\left(\frac{x_\perp}{r^2}\right)_0\delta(x^{-})(x^{-\prime})^2$, which can be integrated under the normalization $\int_{\sigma_1}^{\sigma_2} d\sigma x_\perp^\prime=-\bar{x}_\perp$ to give
\begin{align}
 x_\perp(\sigma)^\prime=-\frac{\bar{x}_\perp}{\Delta_{21}}+\frac{1}{2}(\partial_\perp f)_0x^{-\prime}\left\{ \theta(x^-(\sigma))-\langle \theta\rangle \right\}, \label{solution}
\end{align}
where $\langle \theta \rangle=\int_{\sigma_1}^{\sigma_2}\frac{d\sigma^\prime}{\Delta_{21}}\theta(x^{-})$, and $(\cdots)_0$ means the value at $\sigma=\sigma_0$ satisfying $x^{-}(\sigma_0)=0$. Then it follows that
\begin{align}
 x_\perp^{\prime 2}=-\frac{1}{\Delta_{21}}x_\perp^\prime\cdot \bar{x}_\perp+\frac{1}{2}(f^\prime)_0 x^{-\prime}(\theta-\langle\theta\rangle)+\frac{1}{2}\left\{x_\perp^\prime-(x_\perp^\prime)_0\right\}\cdot \left(\partial_\perp f \right)_0 x^{-\prime}(\theta-\langle\theta\rangle). \label{x-perp}
\end{align}
Since one can also write $x_\perp^\prime-(x_\perp^\prime)_0=Q\left(\frac{x_\perp}{r^2}\right)_0\frac{\bar{x}^{-}}{\Delta_{21}}\{\theta(x^{-})-\theta(0)\}$ again from equation (\ref{solution}), the integral of (\ref{x-perp}) in the interval $(\sigma_1,\sigma_2)$ gives
\begin{align}
 \frac{\Delta_{21}}{2}\int_{\gamma,\sigma_1}^{\sigma_2}d\sigma x_\perp^{\prime 2}=\frac{1}{2}\bar{x}_\perp^2-\frac{1}{2}\left(\frac{Q}{r}\right)_0^2 \bar{x}^{-}\int_{\sigma_1}^{\sigma_2}d\sigma x^{-\prime}\theta(x^{-})\left\{\theta(x^{-})-\langle\theta\rangle \right\}\simeq \frac{1}{2}\bar{x}_\perp^2
\end{align}
by disregarding the $Q^2(\sim 1/E_P^2)$ term as negligible small quantity.  

On the other hand, the $\delta(x^{-})$ term in the right-hand side of (\ref{geodesic-2}) remains providing $x^{-}(\sigma_0)(=0)$ is wedged in between $x^{-}_{(1)}$ and $x^{-}_{(2)}$; then,
\begin{align}
 \frac{\Delta_{21}}{2}\int_{\sigma_1}^{\sigma_2}d\sigma f(x_\perp)\delta(x^{-})(x^{-\prime})^2=-\frac{\bar{x}^{-}}{2}(f)_0\left[ \theta(x^{-})\right]_{1}^{2}.
\end{align}
If it is necessary, further, we can use the form
\footnote{For example for, $x^{-}_{(2)}>0$ and $x^{-}_{(1)}<0$, the difference of the integrals of (\ref{solution}) in the interval $(\sigma_2,\sigma_0)$ and $(\sigma_0,\sigma_1)$ gives rise to $x_\perp(\sigma_0)=X_\perp+\frac{\bar{x}_\perp}{\Delta_{21}}\left(\frac{\sigma_1+\sigma_2}{2}-\sigma_0\right)-(\partial_\perp f)_0\frac{x^{-\prime}}{\Delta_{21}}(\sigma_2-\sigma_0)(\sigma_0-\sigma_1)$. Further, $x^{-}(\sigma_0)=0$ gives $\frac{1}{\Delta_{21}}\left(\frac{\sigma_1+\sigma_2}{2}-\sigma_0\right)=-\frac{X^{-}}{\bar{x}^{-}}$; and, we obtain $x_\perp(\sigma_0)=X_\perp-\bar{x}_\perp\frac{X^{-}}{\bar{x}^{-}}+\mbox{O}(Q)$.}
$(f)_0\simeq f\left(X_\perp-\bar{x}_\perp\frac{X^{-}}{\bar{x}^{-}}\right)+\mbox{O}(Q^2)$ .
Therefore, the interaction terms defined from the geodesic interval can be represented as
\begin{align}
 V(x_{(1)},x_{(2)})\simeq \kappa^2 \left(\bar{x}^2-\bar{x}^{-}(f)_0\left[\theta(x^{-})\right]_{1}^{2} \right) +\omega
\end{align}
disregarding the terms of the order of $\mbox{O}(Q^2)$.

\subsection{Canonical formulation of bi-local model}

Varying the action (\ref{action}) with respect to $e_{(i)},(i=1,2)$, one can obtain the constraints, which are characteristic of the bi-local model, such that
\begin{align}
 H_i\equiv g^{\mu\nu}p_{(i)\mu} p_{(i)\nu}+V=-2p_{(i)+}p_{(i)-}+p_{(i)\perp}^2-f(x_{(i)\perp})\delta(x_{(i)}^{-})p_{(i)+}^2+V=0, \label{H_i=0}
\end{align}
where $p_{(i)\mu}=\frac{\delta S}{\delta x_{(i)}^\mu}=\frac{1}{e_{(i)}}g_{\mu\nu}\dot{x}_{(i)}^\nu,(i=1,2)$ are momenta conjugate to $x_{(i)}^\mu,(i=1,2)$. The $\delta$-function type of singularities in $H_i$ can be removed by the canonical transformation $\tilde{H}_i=UH_iU^\dag,(i=1,2)$ with
\footnote{
In c-number theory, $e^{-iK}L(e^{-iK})^\dag \equiv \sum_n\frac{1}{n!}(\mbox{ad}\,K)^n L$ with $(\mbox{ad}\,K)L=\{K,L\}$, where $\{*,*\}$ is the Poisson bracket. In q-number theory, $U$ itself is a unitary transformation.
}
\begin{align}
 U=\exp\left\{-\frac{i}{2}\sum_{i=1}^2 f(x_{(i)\perp})\theta(x_{(i)}^{-})p_{(i)}^{-}\right\}, \label{unitary transformation}
\end{align}
from which we obtain 
\begin{align}
U
\begin{Bmatrix}
 {p}_{(i)}^{+} \\
 {p}_{(i)}^{-} \\
 {p}_{(i)\perp}
\end{Bmatrix}
U^\dag &=
\begin{Bmatrix}
 \tilde{p}_{(i)}^{+} \\
\tilde{p}_{(i)}^{-} \\
 \tilde{p}_{(i)\perp}
\end{Bmatrix}
=
\begin{Bmatrix}
p_{(i)}^{+}-\frac{1}{2}f(x_{(i)\perp})\delta(x_{(i)}^{-})p_{(i)}^{-} \\
p_{(i)}^{-} \\
p_{(i)\perp}-A(x_{(i)\perp})\theta(x^{-}_{(i)})p^{-}_{(i)}
\end{Bmatrix},
\\ \\
U
\begin{Bmatrix}
 {x}_{(i)}^{-} \\
 {x}_{(i)}^{+} \\
 {x}_{(i)\perp}
\end{Bmatrix}
U^\dag &=
\begin{Bmatrix}
 \tilde{x}_{(i)}^{-} \\
 \tilde{x}_{(i)}^{+} \\
 \tilde{x}_{(i)\perp}
\end{Bmatrix}
=
\begin{Bmatrix}
x_{(i)}^{-} \\
x_{(i)}^{+} +\frac{1}{2}f(x_{(i)\perp})\theta(x^{-}_{(i)})\\
x_{(i)\perp}
\end{Bmatrix},
\end{align}
where
\begin{align}
 A(x_\perp)=-\frac{1}{2}\left(\partial_\perp f(x_\perp)\right)=Q\frac{x_\perp}{r^2}.
\end{align}
Then the addition $\tilde{H}=4(\tilde{H}_1+\tilde{H}_2)$ and the subtraction $\tilde{T}=\tilde{H}_1-\tilde{H}_2$ have the expressions
\begin{align}
 \tilde{H} &=4\left(\frac{1}{2}\sum_{i=1}^2\left\{-2p^{+}p^{-}+(p_\perp -A\theta p^{-})^2\right\}_{(i)}+\tilde{V} \right) \nonumber \\
 &=\left\{ P^2+4\left(\bar{p}^2+\tilde{V} \right) \right\}+2\sum_{i=1}^2\left(-\{p_\perp,A \}+A^2p^{-}\right)_{(i)}(\theta p^{-})_{(i)}, \label{tilde-H}\\
 \tilde{T}&=\sum_{i=1}^2(-1)^{i-1}\left\{-2p^{+}p^{-}+(p_\perp -A\theta p^{-})^2\right\}_{(i)} \nonumber \\
 &=2P\cdot\bar{p}+\sum_{i=1}^2(-1)^{i-1}\left(-\{p_\perp,A \}+A^2 p^{-}\right)_{(i)}(\theta p^{-})_{(i)}, \label{tilde-T}
\end{align}
with
\begin{align}
 \tilde{V} =V\left(x^{-}_{(i)},\tilde{x}^{+}_{(i)},x_{\perp(i)}\right)
=\kappa^2 \left(\bar{x}^2-\bar{x}^{-}\left[\{(f)_0-f\}\cdot \theta \right]_{1}^{2} \right) +\omega \simeq \kappa^2\bar{x}^2 +\omega .
\end{align}
Here, in the last equality, we have regarded that the $[\{(f)_0-f\}\theta]^2_1$ term is a higher-order-small quantity than $Q$, since $(f)_0-f\propto Q$ and $[\theta]^2_1\propto \kappa^{-1}$. By the same reasons, we may disregard the subtraction term $\sum_{i=1}^2(-1)^{i-1}\left(\{p_\perp,A\}\theta p^{-}\right)_{(i)}$, which is decomposed into two higher-order-small quantities: $\frac{1}{2}[\theta]^{2}_{1}\sum_i\left(\{p_\perp,A\}p^{-}\right)_{(i)}$ and $\frac{1}{2}\left(\sum_i\theta_{(i)}\right)[\{p_\perp,A\}p^{-}]^{2}_{1}$. In our standpoint, the $A^2(\propto Q^2)$ terms in $\tilde{H}$ and $\tilde{T}$ are of course negligible. Therefore, the practical forms of (\ref{tilde-H}) and (\ref{tilde-T}), which we are dealing with from now on are
\begin{align}
 \tilde{H} &=P^2+4\left(\bar{p}^2+\kappa^2 \bar{x}^2 \right) +\omega +\Delta M^2, \label{wave-eq-1}\\
  \tilde{T} &=P\cdot \bar{p}, \label{subsidiary-1} 
\end{align}
where $P=p_{(1)}+p_{(2)}$, $\bar{p}=\frac{1}{2}(p_{(1)}-p_{(2)})$, and
\begin{align}
 \Delta M^2=-2\sum_{i=1}^2\left(\{p_\perp,A \}\theta p^{-}\right)_{(i)}.
\end{align}

In q-number theory, thus, the constraints (\ref{H_i=0}) are reduced respectively to the wave equation $\tilde{H}|\tilde{\Phi}\rangle=0$ of the bi-local model and its subsidiary  condition $\langle\tilde{\Phi}|\tilde{T}|\tilde{\Phi}\rangle=0$; we use the subsidiary  condition  in the sense of expectation value by taking into account the consistency between two constraints in Minkowski spacetime. Then in terms of the oscillator variables (\ref{oscillator}),  the wave equation and a sufficient form of subsidiary condition can be written as 
\begin{align}
  & \left(P^2+\alpha^{\prime -1}a^\dag \cdot a+m^2+\Delta M^2\right)|\tilde{\Phi}\rangle=0 , \label{wave-eq-2} \\
 & \hspace{10mm} \hat{P}\cdot a|\tilde{\Phi}\rangle=0,~  \left(\, \hat{P}^\mu=P^\mu/\sqrt{-P^2} \,\right) \label{subsidiary-2}
\end{align}
where $\alpha^\prime=\frac{1}{8\kappa}$ and $m_0^2=4\omega+16\kappa$ as in Appendix(A).

In the case of $\Delta M^2=0$, the equations (\ref{wave-eq-2}) and (\ref{subsidiary-2}) are reduced to those of free bi-local model in Minkowski spacetime, which are compatible each other. For $\Delta M^2\neq 0$, the compatibility is spoiled; a simple way to avoid of this problem is to replace $\Delta M^2$ by its physical component $[\Delta M^2]$ defined by $I\Delta M^2 I=[\Delta M^2]I$, where $I(=I^2)$ is the projection operator onto the subspace characterized by (\ref{subsidiary-2}). Another way  is to replace $x_{(i)}^\mu,(i=1,2)$ in $\Delta M^2$ by their physical components $[x_{(i)}^\mu]$
\footnote{
One can write $I=\frac{1}{2\pi}\int_0^{2\pi}d\theta e^{i\theta N}$ and $[L]=\frac{1}{2\pi}\int_0^{2\pi}d\theta e^{i\theta N}Le^{-i\theta N}$ $\left(\, N=(\hat{P}\cdot a^\dag)(\hat{P}\cdot a)\, \right)$, from which 
$
 [x_{(i)}^{\mu}]=\left(X^\mu+\frac{P_\nu S^{\nu\mu}}{P^2} \right)+(-1)^{i-1}\left(\eta^{\mu\nu}-\frac{P^\mu P^\nu}{P^2}\right)\bar{x}_\nu,\, \left(\, S^{\nu\mu}=ia^{\dag[\nu}a^{\mu ]} \,\right)$ is obtained\cite{Naka-Kakuhata}.
}.
In the following, for the sake of simplicity, we use the first approach to extract a physical component out of $\Delta M^2$, although there arises difference in two approaches in some cases\cite{Naka-Kakuhata}D

\section{Scattering of bi-local model in the shock-wave background}

The scattering matrix of the bi-local model by the gravitational-shock-wave background can be evaluated from the states satisfying wave equation (\ref{wave-eq-2}) followed by the unitary transformation (\ref{unitary transformation}). For this purpose, it is convenient to use the light-like time $T^{-}=X^{-}$ and the corresponding representation of conjugate momentum $P^{+}=i\frac{\partial}{\partial T^{-}}$. With this choice of time parameter, the wave equation (\ref{wave-eq-2}) can be written in the form of Schr\"odinger equation: 
\begin{align}
 i\frac{\partial}{\partial T^{-}}|\tilde{\Phi}\rangle =\frac{1}{2\alpha^\prime P^{-}}\left\{ \alpha^\prime(P_\perp^2+m_0^2)+a^\dag\cdot a+\alpha^\prime [\Delta M^2] \right\}|\tilde{\Phi}\rangle
=(\tilde{H}_0+\Delta\tilde{H})|\Phi\rangle,
\end{align}
where
\begin{align}
 \tilde{H}_0 &=\frac{1}{2\alpha^\prime P^{-}}\left\{ \alpha^\prime(P_\perp^2+m_0^2)+a^\dag\cdot a \right\}, \\
 \Delta\tilde{H} &=\frac{1}{2P^{-}}[\Delta M^2].
\end{align}
 In the scattering under consideration, we deal with the case $|\bar{p}^{-}/P^{-}|\simeq 0$; then, it is no problem to approximate $p_{(i)}^{-}\simeq \frac{1}{2}P^{-}$ in addition to $\theta_i\simeq \theta(X^{-})$ due to $[\theta]_1^2\sim\kappa^{-1}$. Then, with help of $\{p_\perp,A\}=-\frac{i}{2}[p_\perp^2,f]$, we represent the $\Delta\tilde{H}$ in such a convenient form for later use as
\begin{align}
 \Delta\tilde{H}=-\frac{1}{2}\left[\sum_i\{p_\perp,A\}_{(i)}\right]\theta(X^{-})=\frac{i}{2}[\tilde{H}_0,f_{12}]P^{-}\theta(X^{-}), \label{Delta-H}
\end{align}
where 
\begin{align}
 f_{12}=\sum_{i=1}^2 f_i=2f_0 -2Q\log\left(\frac{r_1r_2}{r_0^2}\right).
\end{align}
Under those considerations, the time-displacement operator from $T_1^{-}$ to $T_2^{-}$, the $U$-matrix, in $\{\tilde{\Phi} \}$ space is formally given by$\mbox{T}\exp\left\{-i\int_{T_1^{-}}^{T_2^{-}}dT^{-}(\tilde{H}_0+\Delta\tilde{H})\right\}$. Here, $\mbox{T}\exp\{\cdots\}$ stands for the time ordered exponential with respect to $T^{-}$, which is necessary by the presence of $X^{-}$ dependence of $\Delta M^2$. The $S$-matrix in $\{\Phi=U^\dag\tilde{\Phi}\}$ space is, thus, given by
\begin{align}
 S &=\lim_{\tiny
\begin{matrix}T_2^{-}\rightarrow\infty \\
T_1^{-}\rightarrow -\infty
\end{matrix}}
e^{iT_2^{-}{H}_0}U_2^\dag 
e^{-iT_2^{-}\tilde{H}_0}\left(Te^{-i\int_{T_1^{-}}^{T_2^{-}}dT^{-}(\Delta\tilde{H})_D}\right)e^{iT_1^{-}\tilde{H}_0}U_1e^{-iT_1^{-}{H}_0} \\
 &=\lim_{\tiny
\begin{matrix}T_2^{-}\rightarrow\infty \\
T_1^{-}\rightarrow -\infty
\end{matrix}}
U_2^\dag\left(Te^{-i\int_{T_1^{-}}^{T_2^{-}}dT^{-}(\Delta\tilde{H})_D}\right)U_1
\label{S-matrix}
\end{align}
where $(\Delta\tilde{H})_D=e^{i\tilde{H}_0T^{-}}\Delta\tilde{H}e^{-i\tilde{H}_0T^{-}}$ is the Dirac picture of $\Delta\tilde{H}$. 

Now, one can say that $\langle\Phi_b|S-1|\Phi_a\rangle \propto \delta(P_b^{-}-P_a^{-})$ because of $[P^{-},S]=0$; and so,  the $T$-matrix elements among the states $|\Phi_i\rangle=|\phi_i(\bar{x})\rangle\otimes|P_{\perp i}\rangle\otimes |P_i^{-}\rangle\, ,(i=a,b,\cdots)$ are defined by
\begin{align*}
 \langle \Phi_b|S|\Phi_a\rangle=\langle \Phi_b|\Phi_a\rangle +i(2\pi)\delta(P_b^{-}-P_a^{-})T_{ba}.
\end{align*}
By taking $U_1\rightarrow 1\,(T_1^{-}\sim -\infty)$ and $U_2\rightarrow e^{-\frac{i}{2}\sum_i(fp^{-})_{(i)}} \, (T_2^{-}\sim \infty)$ into account, we obtain the expression of $T$-matrix elements at those asymptotic times $T_1^{-},\, T_2^{-}$ such that
\begin{align}
 T_{ba}=-\frac{i}{(2\pi)}\langle P_{\perp b}|\otimes\langle\phi_b|\left[U_2^\dag \left(Te^{-i\int_{T_1^{-}}^{T_2^{-}}dT^{-}(\Delta\tilde{H})_D}\right)-1 \right]|\phi_a\rangle\otimes|P_{\perp a}\rangle, \label{T-matrix}
\end{align}
where
\begin{align}
 U_2^\dag =e^{\frac{i}{2}P^{-}\left\{f_0-Q\log\left(\frac{r_1r_2}{r_0^2}\right)\right\}}e^{-iQ\bar{p}^{-}\log\left(\frac{r_1}{r_2}\right)}, \label{T-matrix-2}
\end{align}
and $r_i=\sqrt{(X_\perp+(-1)^{i-1}\frac{1}{2}\bar{x}_\perp)^2},\,(i=1,2)$. In practice, since $(\Delta\tilde{H})^2$ is a negligible small quantity, it is sufficient to evaluate (\ref{T-matrix}) within the approximation of the first order of $\Delta\tilde{H}$; that is, we may expand $T_{ab}=T_{ab}^{(0)}+T_{ab}^{(1)},\, (T_{ab}^{(n)}\sim O((\Delta\tilde{H})^n))$. Thus, the next task is to evaluate $T_{ab}^{(n)},\,(n=0,1)$ by using (\ref{T-matrix}) and (\ref{T-matrix-2}) for some cases. \vspace{3mm}

\noindent
case (1) \vspace{3mm}

Let us consider the case of $|\phi_a\rangle=|0_{-}\rangle\otimes |0_{\perp} \rangle$ and $|\phi_b\rangle=|\bar{0}_{-}\rangle\otimes |0_\perp \rangle$, where $|0_\perp\rangle$ is defined by $a_\perp|0\rangle=0$ with $\langle 0_\perp|0_\perp \rangle=1$. Further, $|0_{-}\rangle,\,|\bar{0}_{-}\rangle$ are the states characterized by $a^{-}|0_\pm\rangle=a^{-\dag}|0_\pm\rangle=0$ and $\langle\bar{0}_{-}|a^{+}=\langle\bar{0}|a^{+\dag}=0$with $\langle \bar{0}_{-}|0_{-}\rangle=1$. In this case, since $\bar{x}^{-}|0_{-}\rangle =\bar{p}^{-}|0_{-}\rangle =0$, the expression (\ref{Delta-H}) is exact on those states. Then, for $P_{\perp b}\neq P_{\perp a}$, we obtain
\begin{align}
 T_{ba}^{(0)} =\frac{e^{\frac{i}{2}f_0 P^{-}_a}}{(2\pi)i}\int \frac{d^2X_\perp}{(2\pi)^2}e^{-i(P_{\perp b}-P_{\perp a})\cdot X_\perp } \langle 0_\perp|e^{-\frac{i}{2}P^{-}_aQ\log\left(\frac{r_1r_2}{r_0^2}\right)}|0_\perp\rangle
\label{T-0}
\end{align}
with the normalization $\langle X_\perp|P_\perp\rangle=e^{iP_\perp\cdot X_\perp}/(2\pi)$. In the right-hand side of (\ref{T-0}), the center of mass variables $X_\perp$ in $r_i$'s are already c-numbers. Then, the expectation value by the ground state can be calculated as follows (Appendix C):
\begin{align}
\langle 0|e^{-\frac{i}{2}P^{-}_aQ\log\left(\frac{r_1 r_2}{r_0^2}\right)}|0\rangle=e^{-\frac{i}{2}P_a^{-}Q\log\left(\frac{X_\perp^2}{r_0^2}\right)} \left[ 1-i\pi\alpha^\prime P_a^{-}Q\delta^2(X_\perp)+\cdots \right]. \label{expand-1}
\end{align}
In order to pick up the effect of the second term in the right-hand side of this equation, it is necessary to regularize $\log(X_\perp^2/r_0^2)$ in some way. Then the second term adds a correction of the order of $\alpha^\prime P^{-}Q$ to (\ref{T-0}).  In the zero-slope limit $\alpha^\prime\rightarrow 0$, which pick out a point-like limit of the bi-local model, the first term yields the exact result of expectation value. Substituting this first term for (\ref{T-0}), we obtain
\footnote{
According as \cite{tHooft}, $
 \int d^2X_\perp e^{i\Delta P_\perp \cdot X_\perp-iB\log\left(\frac{X_\perp^2}{r_0^2}\right)}=r_0^2\frac{\pi\Gamma(1-iB)}{\Gamma(iB)}\left(\frac{4}{r_0^2\Delta P_\perp^2}\right)^{1-iB}$ has been used.}
\begin{align}
 \lim_{\alpha^\prime\rightarrow 0}T_{ba}^{(0)}=\frac{e^{\frac{i}{2}f_0 P^{-}_a}}{(2\pi)^3i}\times r_0^2\frac{\pi\Gamma(1-\frac{i}{2}P_a^{-}Q)}{\Gamma(\frac{i}{2}P_a^{-}Q)}\left\{\frac{4}{r_0^2(\Delta P_{\perp})^2}\right\}^{1-\frac{i}{2}P_a^{-}Q},(\Delta P_\perp=P_{\perp b}-P_{\perp a}).
\end{align}
The result is nothing but the one of G. 't Hooft except the factor $f_0$, which is simply a result of normalization $f_0=f(r_0)$ in the present case.

When we calculate $T^{(1)}_{ba}$ as the next task, it should be noticed that the equation (\ref{Delta-H}) allows us to write
\begin{align}
 -i\int_{T_1^{-}}^{T_2^{-}}dT^{-}(\Delta\tilde{H})_D=-\frac{i}{2}\int_{T_1^{-}}^{T_2^{-}}dT^{-}\frac{d}{dT^{-}}(f_{12})_DP^{-}\theta(T^{-})
=\frac{i}{2}f_{12}P^{-}
\end{align}
with the substitution $f_{12}\rightarrow f_{12}e^{-\epsilon T^{-}},(\epsilon=+0)$. Then we have
\begin{align}
 T^{(1)}_{ba} &=\frac{1}{2(2\pi)}\langle P_{\perp b}|\otimes \langle 0_\perp|U_2^\dag f_{12}P_a^{-}|0_\perp \rangle\otimes |P_{\perp a}\rangle \nonumber \\
 &=\frac{e^{\frac{i}{2}f_0 P^{-}_a}P_a^{-}}{(2\pi)}\int \frac{d^2X_\perp}{(2\pi)^2}e^{-i(P_{\perp b}-P_{\perp a})\cdot X_\perp } \langle 0_\perp |e^{-\frac{i}{2}P^{-}_aQ\log\left(\frac{r_1r_2}{r_0^2}\right)}\left\{f_0-Q\log\left(\frac{r_1r_2}{r_0^2}\right)\right\}|0_\perp \rangle  \nonumber \\
 &=if_0P_a^{-}T^{(0)}+2iP_a^{-}e^{\frac{i}{2}f_0 P^{-}_a}\frac{\partial}{\partial P_a^{-}}\left(e^{-\frac{i}{2}f_0 P^{-}_a}T^{(0)}\right). \label{T1}
\end{align}
The resultant expression means that the role of $\kappa$ may not be effective in $T_{ba}^{(1)}$ too by the same reason as in $T_{ba}^{(0)}$. \\

\noindent
case (2) \vspace{3mm}

Secondly, let us consider the case of $|\phi_a\rangle=|z_a,\bar{z}_a\rangle\otimes |0_{\perp} \rangle$ and $|\phi_b\rangle=|\bar{0}_{-}\rangle\otimes |0_\perp \rangle$, where $z_a=\sqrt{\frac{\kappa}{2}}\bar{x}_a^{-}+\frac{i}{\sqrt{2\kappa}}\bar{p}_a^{-}$ and $\bar{z}_a=z_a^*=\sqrt{\frac{\kappa}{2}}\bar{x}_a^{-}-\frac{i}{\sqrt{2\kappa}}\bar{p}_a^{-}$. The $|z,\bar{z}\rangle$ is the eigenstate
\footnote{The $|z,\bar{z}\rangle$ is also a coherent state having the form $|z,\bar{z}\rangle=e^{-a^{+\dag}+\bar{z}a^{+}}|0_{-}\rangle$, to which $a^{-}|z,\bar{z}\rangle=z|z,\bar{z}\rangle$ and $a^{-\dag}|z,\bar{z}\rangle=\bar{z}|z,\bar{z}\rangle$ hold. With the adjoint state $\langle z,\bar{z}|=\langle\bar{0}_{-}|e^{-z^*a^{-}+\bar{z}^*a^{-\dag}}$, $I=\int\frac{d^2z}{\pi}e^{-|z|^2}\int\frac{d^2\bar{z}}{\pi}e^{-|\bar{z}|^2}|z,\bar{z}\rangle\langle -z,-\bar{z}|$ becomes the unit operator in this representation.
}
 of $(\bar{x}^{-},\bar{p}^{-})$ with eigenvalues $(\bar{x}_a^{-},\bar{p}_a^{-})$; and so, the bi-local model is consisting of two particles in different times $x_{(i)}^{-},\,(i=1,2)$ and momenta $p_{(i)}^{-},\,(i=1,2)$. Without loss of generality, however, we may put $\bar{x}_a^{-}=0$, since its effect is absorbed by a large $|T^{-}_{a}|\sim \infty
$. Then, using the way evaluating the  expectation value by  $|0_\perp\rangle$ in (Appendix C), (\ref{T-matrix}) comes to be

\begin{align}
 T_{ba}^{(0)}  &=\frac{e^{\frac{i}{2}f_0 P^{-}_a}}{2\pi i}
\langle P_{\perp b}|\otimes \langle \bar{0}_{-}|
e^{-\frac{i}{2}P^{-}Q\log\left(\frac{r_1r_2}{r_0^2}\right)}e^{-iQ\bar{p}^{-}\log\left(\frac{r_1}{r_2}\right)}
|z_a,\bar{z}_a\rangle\otimes |P_{\perp a}\rangle  \nonumber \\
 &=\frac{e^{\frac{i}{2}f_0 P^{-}_a}}{2\pi i}\int \frac{d^2X_\perp}{(2\pi )^2}e^{-i\Delta P_\perp\cdot X_\perp }
\langle 0_\perp|
e^{Q\bar{p}_a^{-}\log\left(\frac{r_1}{r_2}\right)}
e^{-\frac{i}{2}P^{-}_aQ\log\left(\frac{r_1r_2}{r_0^2}\right)}|0_\perp \rangle \nonumber \\
 &=\frac{e^{\frac{i}{2}f_0 P^{-}_a}}{2\pi i}\int \frac{d^2X_\perp}{(2\pi)^2}e^{-i\Delta P_{\perp b}\cdot X_\perp }e^{-\frac{i}{2}P^{-}_aQ\log\left(\frac{X_\perp^2}{r_0^2}\right)}
\left[1-i\pi\alpha^\prime P_a^{-}Q\delta^2(X_\perp)-\frac{\alpha^\prime}{2} \frac{(Q\bar{p}_a^{-})^2}{r^2}+\cdots\right], \label{T0-2}
\end{align}
where $r^2=X_\perp^2$. The resultant form of $T^{(0)}_{ba}$ includes a correction term of the order of $\alpha^\prime(Q\bar{p}^{-})^2$ in addition to the $\delta$-function type of correction. The term is seemingly negligible; for a case such as $|\bar{p}_a^{-}|\gtrsim Q^{-\frac{1}{2}}$, however, this term comes to be comparable order to the $\delta$-function term.  As for $T_{ba}^{(1)}$  (\ref{Delta-H}) is again available to use though it is an approximate equation in this case; that is, the $T_{ba}^{(1)}$ is calculated according as (\ref{T1}) from (\ref{T0-2}). In the above calculation, we may exchange the role of $|\phi_a\rangle$ for $|\phi_b\rangle$; then, there causes substitution $\bar{p}_a^{-}\rightarrow \bar{p}_b^{-}$ in (\ref{T0-2}).  \\\

\noindent
case (3) \vspace{3mm}

Lastly, we consider the case of $|\phi_a\rangle=|0\rangle\otimes |0_\perp\rangle$  and $|\phi_b\rangle=|\bar{0}\rangle\otimes |\bar{x}_{\perp b}\rangle$, where $|x_{\perp b}\rangle$'s are characterized by $\bar{x}_\perp|\bar{x}_{\perp b}\rangle=\bar{x}_{\perp b}|\bar{x}_{\perp b}\rangle$ and $\langle \bar{x}_{\perp b}|\bar{x}_{\perp c}\rangle=\delta^2(\bar{x}_{\perp b}-\bar{x}_{\perp c})$. In this case, the final states are the superposition of mass eigenstates, to which one can verify $|x_{\perp b}\rangle=\sqrt{\frac{\kappa}{\pi}}e^{-\frac{1}{2}\kappa \bar{x}_{\perp b}^2}e^{-\frac{1}{2}a_\perp^{\dag 2}+\bar{x}_{\perp b}\sqrt{2\kappa}\cdot a_\perp^\dag}|0\rangle$. Then the $T_{ba}^{(0)}$ becomes

\begin{align}
 T_{ba}^{(0)} &=e^{\frac{i}{2}f_0 P^{-}_a}\int \frac{d^2X_\perp}{2\pi i}\frac{e^{-i\Delta_{ba}P_\perp\cdot X_\perp }}{(2\pi)^2} \langle \bar{x}_{\perp b}|e^{-\frac{i}{2}P_a^{-}Q\log\left(\frac{r_1r_2}{r_0^2}\right)_b}|0_\perp\rangle \nonumber \\
 &=e^{\frac{i}{2}f_0 P^{-}_a}\int \frac{d^2X_\perp}{2\pi i}\frac{e^{-i\Delta_{ba}P_\perp\cdot X_\perp }}{(2\pi)^2} \sqrt{\frac{\kappa}{\pi}}e^{-\frac{1}{2}\kappa \bar{x}_{\perp b}^2} e^{-\frac{i}{2}P^{-}_aQ\log\left(\frac{r_1r_2}{r_0^2}\right)_b}.
\end{align}
If we integrate $T_{ba}^{(0)}$ with respect to $\bar{x}_\perp$, the result describes a scattering for the final states consisting of all eigenstates of $\bar{x}_\perp$. Since the integral can be evaluated approximately by means of the saddle point method as

\begin{align} 
 \int d^2\bar{x}_{\perp b}e^{-\frac{1}{2}\kappa \bar{x}_{\perp b}^2} e^{-\frac{i}{2}P^{-}_aQ\log\left(\frac{r_1r_2}{r_0^2}\right)_b}
  \simeq e^{-\frac{i}{2}P^{-}_aQ\log\left(\frac{X_\perp^2}{r_0^2}\right)}\frac{2\pi\kappa}{1+\frac{2i\alpha^\prime QP_a^{-}}{X_\perp^2}},
\end{align}
the superposition of $T_{ba}^{(0)}$ by the integral with respect to $\bar{x}_{\perp b}$ gives
\begin{align}
 T_{*a}^{(0)}=\sqrt{\frac{\kappa}{\pi}}\int d^2\bar{x}_{\perp b}T_{ba}^{(0)} \simeq 
e^{\frac{i}{2}f_0 P^{-}_a}\int \frac{d^2X_\perp}{\pi i}\frac{e^{-i\Delta P_\perp\cdot X_\perp }}{(2\pi)^2}
\frac{e^{-\frac{i}{2}P^{-}_aQ\log\left(\frac{X_\perp^2}{r_0^2}\right)}}
{1+\frac{2i\alpha^\prime QP_a^{-}}{X_\perp^2}}~~.
\end{align}
Since we may write $(1+2i\alpha^\prime QP_a^{-}/X_\perp^2)^{-1}\simeq 1-2i\alpha^\prime QP_a^{-}/X_\perp^2$, we can see that the correction for the $T$-matrix in this case becomes the order of $O(\alpha^\prime QP_a^{-})$.

\section{Summary and discussion}

In this paper, we have discussed the interaction between a bi-local model, the two-particle system bounded by a harmonic-oscillator type of potential, and the shock-wave gravity generated by a point particle with Planck scale momentum. A purpose of this paper is to study the bound-state effects in this interaction.

In formulating the bi-local model interacting with such a gravitational field, we first studied the embedding of the bi-local model within the corresponding curved spacetime. Since the metric representing the shock-wave gravity contains $\delta$-function type of singularity, the action of the bi-local model and the constraints derived from that also contain such a singularity. In the stage of canonical formalism, however, we could remove those singularities by carrying out a canonical transformation in addition to an approximation neglecting the quantities of the order of $O(1/E_P^2)$. As a result, the interaction of bi-local model is reduced to a potential-scattering problem followed by the canonical transformations for  incoming and outgoing states. Then, it is shown that the effect of the shock-wave gravity arises from both of the interaction term $\Delta\tilde{H}$ after canonical transformations and unitary operators $U_i$ of the canonical transformations. In other words, in our formalism, the scattering matrix is not trivial even in the lowest order of $\Delta\tilde{H}$.

Under this setup of the interaction between the bi-local model and S.W.G., we have studied the scattering of the bi-local model in three cases. The first  case is that both of incoming and outgoing states are the ground state of the bi-local model. Then, a bound-state effect appears as a term being proportional to $\alpha^\prime P^{-}Q,\,(Q\sim O(1/E_P))$ in the lowest order $T$-matrix $T^{(0)}_{ba}$. Furthermore, it is confirmed that the zero-slope limit just identical to the result obtained by {}'t Hooft. It is also able to show that the first order correction of $T$-matrix, the $T^{(1)}_{ba}$, is obtained from $T^{(0)}_{ba}$ by a simple manipulation.

Secondary, we studied the case of outgoing states having the structure of coherent states of light-like oscillation with respect to relative coordinates. In this case, the $T^{(0)}_{ba}$ contains additional term being proportional to $\alpha^\prime(Q\bar{p}^{-})^2$ to the previous case. Then, it is not necessary to regard this term so as to be negligible, since for $|\bar{p}^{-}|\gtrsim Q^{-\frac{1}{2}}$, this term comes to be comparable order of $\alpha^\prime P^{-}Q$. Lastly, the study is made on the case such that the outgoing states are eigenstates of $\bar{x}_\perp$, the transverse components of relative coordinates. The superpose of $T$-matrix in this case with respect to various $\bar{x}_\perp$, again gives a quantity of the order of $\alpha^\prime QP^{-}$. Consequently, one can say that the typical order of corrections to $T$-matrix in the bi-local model for the one in point-like particles is  $\alpha^\prime QP^{-}$.

The order of $\alpha^\prime QP^{-}$ may not be always very small when we can realize the situations such as $P^{-}\sim Q^{-1}$ or $\alpha^\prime P^{-}\sim (E_P/E)$, where the $E(\ll E_P)$ is a typical energy scale of the incoming bi-local model. The second case should be an infinite-slope limit of the bi-local model, in which higher-spin states of the bi-local model degenerate into a massless state. Throughout this paper, however, we have assumed that the $\alpha^\prime$ is a small quantity allowing the zero-slope limit. If we want to evaluate the scattering amplitude allowing the infinite-slope limit, then different ways of approximation will be need in each step of calculations. The attention should also be paid on the physical interaction term $[\Delta\tilde{H}]$ defined out of $\Delta\tilde{H}$.  As remarked in section 2, the way extracting physical component of $\Delta\tilde{H}$ is not unique; and the results may depend on the choice of the ways of extraction in general. To find a rational way for this ends will come to be important in the calculation of higher-order amplitudes. Those are interesting future problems to study.

\section*{Acknowledgments}
The authors wish to thank the members of the theoretical group in Nihon University for their interest in this work and comments.

\appendix

\section{Bi-local fields in Minkowski spacetime}

To make clear the bi-local model under consideration, we briefly summarize the model in Minkowski spacetime. The action of this model is usually set as
\begin{align}
S &=\int d\tau \frac{1}{2}\sum_{i=1}^2 \left\{ e_{(i)}^{-1}\eta_{\mu\nu}\dot{x}_{(i)}^\mu \dot{x}_{(i)}^\nu-V_0\left(x_{(1)},x_{(2)}\right)e_{(i)} \right\}, 
\end{align}
where $x_{(i)}^\mu,~(i=1,2)$ and $V_0$ represent respectively the coordinates of two particles and the interaction between them. Further $e_{(i)}(\tau),(i=1,2)$ are einbeins, which guaranty the invariance of $S$ under the $\tau$ reparametrization. With the aid of this invariance, we can describe the interaction so that the action at a distance happens at the same time-ordering parameter $\tau$. Varying the action with respect to $e_{(i)}$, we obtain the constraints
\begin{align}
 H_i\equiv p_{(i)}^2+V_0=0,~(i=1,2), \label{constraint-1}
\end{align}
where $p_{(i)\mu}=\frac{\delta S}{\delta x_{(i)}^\mu}=\frac{1}{e_{(i)}}\dot{x}_{(i)\mu},(i=1,2)$ are momenta conjugate to $x_{(i)}^\mu$'s. In terms of center of mass momentum $P=p_{(1)}+p_{(2)}$ and relative momentum $\bar{p}=\frac{1}{2}(p_{(1)}-p_{(2)})$, the recombination of the constraints (\ref{constraint-1})  gives
\begin{align}
 \frac{1}{4}H &\equiv\frac{1}{2}(H_1+H_2)=\frac{1}{4}P^2+\bar{p}^2+V_0=0 , \\
 T &\equiv\frac{1}{2}(H_1-H_2)=P\cdot\bar{p}=0 . \label{constraint-2}
\end{align}

In q-number theory, the $H|\Phi\rangle=0$ is the wave equation of the bi-local system, to which it is usually assumed to get a linear $J \propto E^2$ relation that $V_0(x_{(1)},x_{(2)})=\kappa^2\bar{x}^2+\omega\,, (\kappa,\omega=\mbox{const.})$, where $\bar{x}^\mu=x_{(1)}^\mu -x_{(2)}^\mu$ are relative coordinates of the system.  Introducing, here, the oscillator variables $(a^\mu,a^{\mu\dag})$ defined by
\begin{align}
 \bar{x}^\mu=\sqrt{\frac{1}{2\kappa}}\left(a^{\mu\dag}+a^\mu\right),~\bar{p}^\mu=i\sqrt{\frac{\kappa}{2}}\left(a^{\mu\dag}-a^\mu \right), \label{oscillator}
\end{align}
the wave equation becomes
\begin{align}
 H|\Phi\rangle=\left(P^2+\frac{1}{\alpha^\prime}a^\dag\cdot a+m_0^2\right)|\Phi\rangle=0, \label{condition-1}
\end{align}
where $\alpha^\prime=\frac{1}{8\kappa}$ and $m_0^2=4\omega+16\kappa$. Then the commutation relation $[a_\mu,a_\nu^\dag]=\eta_{\mu\nu}$ means that time-like excitations $\{(a_0^\dag)^n|0\rangle\}$ are ghost states. The subsidiary condition (\ref{constraint-2}) has a structure to eliminate those states; however, since $[H,T]=-2i\kappa P\cdot\bar{x}$, (\ref{constraint-1}) and (\ref{constraint-2}) are not compatible. One practical way to get rid of this problem is to use (\ref{constraint-2}) in the sense of expectation value $\langle\Phi|T|\Phi\rangle=0$ as Gupta-Bleuler formalism in Q.E.D.; that is, we put
\begin{align}
 T^{(+)}|\Phi\rangle \equiv P\cdot a|\Phi\rangle=0 \label{condition-2}
\end{align}
as the subsidiary condition that is compatible with (\ref{condition-1}). The condition (\ref{condition-2}) is available to use for the free bi-local model with $V_0=\kappa^2\bar{x}^2+\omega$. In the other cases, we have to deal with the interaction terms under some kind of projections to physical states as done in section 3.

\section{Aichelburg-Sexl boost}

In a curved spacetime with metric $g_{\mu\nu}$, the Lagrangian density of a point particle is 
\begin{align}
 {\mathcal L}=-Mc\int d\tau \frac{ds}{d\tau}\delta^4(x-x(\tau)),~(ds=\sqrt{-dx^\mu g_{\mu\nu}dx^\nu}),
\end{align}
where $M$ is the rest mass of the particle, and $\tau$ is an appropriate time-ordering parameter. The energy momentum density for the particle in the Minkowski spacetime is given by
\begin{align}
 {\mathcal T}^{\mu\nu} &=\left. 2\frac{\delta}{\delta g_{\mu\nu}(x(\tau))}{\mathcal L}\right|_{g_{\mu\nu}=\eta_{\mu\nu}}=Mc\frac{dx^\mu}{ds}\frac{dx^\nu}{d\tau}\left|\frac{d\tau}{dx^0}\right|\delta^3(\bm{x}-\bm{x}(t))  \label{E.M.T 1}
\end{align}
with $ds=cdt\sqrt{1-\beta^2},(\beta=|\frac{d\bm{x}}{dx^0}|)$. Then, under the boost $x^3(t)=\beta x^0$ and $x^1(t)=x^2(t)=0$, the energy momentum tensor density (\ref{E.M.T 1}) with $d\tau=dx^0$ becomes
\begin{align}
 {\mathcal T}^{\mu\nu}=\frac{E_S}{c}u(\beta)^\mu u(\beta)^\nu\delta(x^3-\beta x^0)\delta(x^1)\delta(x^2),~\left( Mc^2=E_S\sqrt{1-\beta^2} \right),
\end{align}
where $E_S=Mc^2/\sqrt{1-\beta^2}$ and $u^\mu=\frac{dx^\mu}{dx^0}$.
The effective massless limit $M \rightarrow 0$ can be realized by taking the limit $\beta \rightarrow 1$ with fixed $E_S$. Then $E_S=cP_S$; and, the resultant energy momentum tensor density becomes
\begin{align}
 {\mathcal T}^{\mu\nu}&=\sqrt{2}P_Se_{+}^\mu e_{+}^\nu \delta(x^{-})\delta^2(x^i),~\left(e_{\pm}^\mu=\frac{1}{\sqrt{2}}(e_0^\mu \pm e_3^\mu) \right).
\end{align}
 \\
The non-zero components of $\mathcal{T}^{\mu\nu}$ is, thus, ${\mathcal T}_{--}=e_{-\mu} {\mathcal T}^{\mu\nu}e_{-\nu}=\sqrt{2}P_S\delta(x^{-})\delta^2(x^i)$ only. Then, it is not difficult to verify that the Einstein equation of the spacetime with the metric $ds^2=-2dx^{+}dx^{-}+F(x^{-},x_\perp)(dx^{-})^2+(dx_\perp)^2$ having $T_{--}=c{\mathcal T}_{--}$ as it source will be reduced to
\begin{align}
 F(x^{-},x^i) &=-\frac{16\pi G}{c^4}\Delta^{-1}{T}_{--}=-\frac{16\pi G}{c^4}\sqrt{2}cP_S\delta(x^{-})\Delta^{-1}\delta^2(x^i),
\end{align}
where $\Delta=\partial_\perp^2$. Therefore, taking $\Delta \log r=2\pi\delta^2(x_\perp)$ into account, $F(x^{-},x_\perp)$ can be solved as $F(x^{-},x_\perp)=\delta(x^{-})f(x_\perp)$ with
\begin{align}
 f(x_\perp)=f_0-2Q\log\left(\frac{r}{r_0}\right)~,~~\left( \, Q=\frac{4\sqrt{2}\hbar c P_S}{E_P^2},\,E_P=\sqrt{\frac{\hbar c^5}{G}} \, \right). \label{Q}
\end{align}

\section{Useful formula}

Taking $\langle \bar{x}_\perp|0\rangle=\sqrt{\frac{\kappa}{\pi}}e^{-\frac{1}{2}\kappa\bar{x}_\perp^2}$ into account, one can write
\begin{align}
 \langle 0|e^{-\frac{i}{2}P_a^{-}Q\log\left(\frac{r_1r_2}{r_0^2}\right)}|0\rangle =\int d^2\bar{x}_\perp f(X_\perp,\bar{x}_\perp)\left(\frac{\kappa}{\pi}\right)e^{-\kappa\bar{x}_\perp^2}, \label{C1}
\end{align}
where the $f(X_\perp,\bar{x}_\perp)$ is the c-number function of $e^{-\frac{i}{2}P_a^{-}Q\log\left(\frac{r_1r_2}{r_0^2}\right)}$; that is,
\begin{align}
 \left[e^{-\frac{i}{2}P_a^{-}Q\log\left(\frac{r_1r_2}{r_0^2}\right)}\right]_{\mbox{c-number}}=f(X_\perp,\bar{x}_\perp) =\int d^2k \tilde{f}(X_\perp,k)e^{ik\bar{x}_\perp}. \label{C2}
\end{align}
Substituting the Fourier integral form (\ref{C2}) for the right-hand side of (\ref{C1}), we obtain
\begin{align*}
 \mbox{r.h.s of (\ref{C1})} &=\int d^2 k \tilde{f}(X_\perp,k)\left(\frac{\kappa}{\pi}\right)\int d^2\bar{x}_\perp e^{-\kappa\bar{x}_\perp^2+ik\bar{x}_\perp}=\int d^2k \tilde{f}(X_\perp,k)e^{-\frac{1}{4\kappa}k^2} \\
  &=e^{-\frac{1}{4\kappa}(-i\partial_y)^2} \int d^2k \left. \tilde{f}(X_\perp,k) e^{iky}\right|_{y=0} \\
 &=e^{-\frac{1}{4\kappa}(-i\partial_y)^2} \left. e^{-\frac{i}{2}P_a^{-}Q\log\left(\frac{r_1(y)r_2(y)}{r_0^2}\right)}\right|_{y=0},
\end{align*}
where $r_1(y)=\sqrt{(X_\perp+\frac{1}{2}y)^2}$ and $r_2(y)=\sqrt{(X_\perp-\frac{1}{2}y)^2}$. Therefore, under the expansion $e^{-\frac{1}{4\kappa}(-i\partial_y)^2} =1+\frac{1}{4\kappa}\Delta_y^2+\cdots$ in addition to $\Delta \log\left(\frac{r}{r_0}\right)=2\pi \delta^2(x_\perp)$, we obtain
\begin{align}
 \langle 0|e^{-\frac{i}{2}P_a^{-}Q\log\left(\frac{r_1r_2}{r_0^2}\right)}|0\rangle =e^{-\frac{i}{2}P^{-}_aQ\log\left(\frac{X_\perp^2}{r_0^2}\right)}\left[1-i\pi\alpha^\prime P_a^{-}Q\delta^2(X_\perp)\cdots\right].
\end{align}

\end{document}